\documentclass[10pt]{elsart}

\usepackage{graphicx}
\usepackage{epsf}

\begin{document}

{

\begin{frontmatter}

\title
{The TeV spectrum of Mkn 501 as measured during the high state in 1997 
by the HEGRA stereoscopic system of imaging air \v{C}erenkov telescopes}

\author{A. Konopelko,}
\author{for the HEGRA collaboration}

\address
{Max-Planck-Institut f\"ur Kernphysik, D-69029 
Heidelberg, Germany}

\begin{abstract}
{The BL Lac object Mkn 501 has shown very high emission in TeV $\gamma$-rays 
from March to October, 1997. During this period the source was continuously 
monitored with the HEGRA stereoscopic system of 4 imaging air \v{C}erenkov 
telescopes for a total exposure time of 110 hours. The unprecedented 
statistics of about 38,000 TeV photons, combined with the good energy resolution 
of $\sim 20$\% over the entire energy range and with detailed simulations of 
the detector response, allowed a determination of the average energy spectrum 
from 500 GeV to 24 TeV. \\
Although the $\gamma$-ray flux varied strongly with time, the daily energy spectra 
remained rather constant in their shape. Therefore it is justified to derive a 
time averaged spectrum and thus to extend, for the first time, the spectral 
measurement well beyond 10 TeV. This TeV spectrum of Mkn 501 shows a gradual 
steepening which could be caused by a number of physical processes, such as a 
limited energy range of the radiating particles, intrinsic $\gamma$-ray 
absorption inside the source, the Klein-Nishina effect (in the case of an 
inverse Compton origin of the radiation) and finally, an absorption of the TeV 
$\gamma$-rays propagating in the intergalactic medium.}
\end{abstract}

\end{frontmatter}

\noindent
{\bf 1. Experiment}

The VHE $\gamma$-ray observatory of the HEGRA collaboration consists of six imaging 
air \v{C}erenkov telescopes (IACTs) located at the Roque de los Muchachos (2.2 km 
height) on the Canary Island of La Palma. The prototype telescope with a mirror 
area of $\rm 5 \, m^2$ effectively started  its operation in 1992 (Mirzoyan et al., 
1994) and since then has undergone several hardware upgrades (Lorenz et al., 1998). 
Nowadays this telescope is used as an independent instrument. The system of 4 IACTs 
has been taking data since 1996 (Daum et al., 1997; Aharonian et al., 1998a) and the 
complete system of 5 IACTs was put into operation in September, 1998. Five identical 
telescopes are set at the corners and center of a quadrangular lot of 100x100 
$\rm m^2$ size. Four telescopes are separated by around 70 m from the central telescope. 
Observations of Mkn 501 have been performed by 4 IACTs in a system. 
Each of telescopes consists of a 8.5 $\rm m^2$ reflector focussing the \v{C}erenkov light onto a 
photomultiplier tube camera. The number of photomultipliers in the camera is 271, 
arranged in a hexagonal matrix covering a field of view with a radius of 
$2.3^\circ$. The angular size of one pixel is $0.25^\circ$. Any telescope camera is 
triggered when the signal in two next-neighbours of the 271 photomultiplier tubes exceeds a threshold 
of 10 (8, since June'97) photoelectrons, and the system readout starts when at least two telescopes were 
triggered by \v{C}erenkov light from an air shower. Currently the array of 5 IACTs gives 
a cosmic ray detection rate of about 14 Hz, which is roughly 1.4 times higher than the 
rate for the 4 IACT system. \\
The IACT system detects at zenith the $\gamma$-ray air showers at impact distances as far as 
250 m from the central telescope, yielding a collection area of $\rm 2\cdot 10^9\,\, cm^2$. 
In the present analysis the collection area, as a function of energy and zenith angle, 
for $\gamma$-ray showers, has been inferred from detailed Monte Carlo simulations 
(Konopelko et al., 1998a). The energy threshold of the telescope system, defined as the 
energy at which the $\gamma$-ray detection rate reaches its maximum for the differential 
spectrum $\rm dN_{\gamma}/dE \sim E^{-2.5}$, is $\sim 0.5$ TeV at small zenith angles, 
and increases up to $\sim2$ TeV at $50^\circ$ zenith angle. The sensitivity of the HEGRA 
IACT system was evaluated by simultaneous observations of the Crab Nebula which is a 
standard candle shining in TeV $\gamma$-rays at an almost constant flux level 
$\rm J_\gamma(>\,1\,TeV) \simeq 1.7 \pm 0.5 \cdot 10^{-11} \,\, \,photons \, cm^{-2} \, s^{-1}$,  
as measured by the HEGRA IACT system (Konopelko et al., 1998b). The measured Crab flux is 
consistent within 20 \% with other measurements performed by several groups using  
ground based imaging air \v{C}erenkov telescopes (Weekes et al., 1997). At present, for 
both the online (the preanalysis routines are installed on the observation site in order 
to control the source activity) and the offline data analysis (the detailed analysis 
taking into account the complete {\it a posteriori} information about the system 
performance), there are two basic options: a search mode and the energy spectrum study 
mode. In the search mode the analysis technique reduces at most the cosmic ray events 
($\kappa_{CR} \leq 10^{-3}$) using {\it tight} orientation and shape cuts still the 
loss of the $\gamma$-rays is of $\sim 50$ \%. For the second option we use a set of  
{\it loose} cuts in order to enrich the $\gamma$-ray statistics ($\kappa_{\gamma} \geq 90$ \%),  
even though the cosmic ray rejection is much lower ($\kappa_{CR} \simeq 10^{-2}$). \\
Note that the stand alone system telescope shows substantially lower 
sensitivity, compared with the system. Thus the test observations of the Crab Nebula in 
1996 with the single system telescope provided $\rm \sim 2.7 \sigma/hr$ detection
and a $\gamma$-ray rate of $\rm \sim 27 \, \gamma/hr$. Assuming that all 4 IACTs are 
operating independently, one can achieve $\rm \sim 5.4 \sigma/hr$ whereas the stereoscopic 
observations yield $\rm 12 \sigma/hr$, at least, using several images from an  
individual air shower. In addition the stereoscopic observations reduce the energy 
threshold from $\sim$800 GeV for a single telescope down to $\sim 500$ GeV. 
Interestingly, the stereoscopic observations give after the loose cuts a $\gamma$-ray 
detection rate which is almost the same as in the case of independent operation of the 
telescopes $\rm R_\gamma = 106 \simeq 27x4 = 108 \gamma/hr$. Thus a $\gamma$-ray source of 
0.25 Crab flux can be detected (at the 3$\sigma$ level) within one hour of observations. 
Preliminary data analysis shows that the complete array of 5 IACTs allows to reach the 
14$\rm \sigma/per \,hour$ detection of the Crab Nebula with a $\gamma$-ray rate of 
$\rm \geq 120 \gamma /hr$. Earlier we have developed the analysis cuts (Daum et al., 1997) 
which are very strong in the background rejection, reducing the cosmic ray rate down to 
$\rm \sim 1 \, hr^{-1}$, though the remaining $\gamma$-ray rate was 24 $\rm \gamma /hr$ 
in Crab Nebular observations. These cuts are of particular interest when searching 
for ``burst like'' $\gamma$-ray signals on very short time scales. 

\noindent
{\bf 2. Observations}

In the present analysis we used only the data runs with 4 operational IACTs. For the 
selected runs the event detection rate varies within about {15 \%} the average rate at 
the relevant zenith angle. The mean angular {\it Width} of the cosmic ray as well as the 
$\gamma$-ray images in each data run deviate from the Monte Carlo predicted value by less 
than 6 \%. Note that the distribution of the image {\it Width} 
(e.g., see Reynolds et al., 1993) for the $\gamma$-ray air showers is very sensitive to the 
improper adjustment of the telescope mirrors, the camera bending etc. The measured 
cosmic ray detection rate has been reproduced by the Monte Carlo simulations with an 
accuracy better than 10 \% utilizing an advanced detector simulation routine. \\
Mkn 501 was observed in a wobble mode; i.e., the telescopes were pointed in 
Declination $\pm 0.^\circ 5$ aside from the nominal Mkn 501 position (the sign of angular 
shift was altered from one run of 20 min to the next). This is useful for continuous 
monitoring of the cosmic-ray background taking the OFF-source region to be symmetric 
about the camera center, offset $1^\circ$ from the ON-source region. This 
simple approach gains a factor of 2 in observation time. Observations were made 
from March 16 to October 1, 1997, for a total of $\sim 110$ hr of data taken at zenith 
angles up to 60 degrees. Note that most of the data were taken at 
small zenith angles up to 30 degree (observation time of $\sim 80$ hr). \\

\noindent
{\bf 3. Analysis}

The stereoscopic imaging analysis of the data is based on the geometrical reconstruction 
of the shower arrival direction and the shower core position in the observation plane, as 
well as on the joint parametrization of the shape of the \v{C}erenkov light images. The 
first $\gamma$-ray selection criterion used here is $\rm \Theta^2\leq 0.05\, \, deg^2$, 
where $\Theta^2$ is the squared angular distance of the reconstructed shower arrival 
direction to the source position. This orientational cut is very loose and gives 90\% 
$\gamma$-ray acceptance whereas the number of cosmic ray showers is reduced by a factor 
of about 50. Further reduction of the cut on $\Theta^2$ was not made since it introduces 
a strong energy dependence of the $\gamma$-ray acceptance. In addition we analysed the 
data by the {\it mean scaled width} parameter,$<\tilde{w}>$. This parameter was introduced 
in order to provide an almost constant $\gamma$-ray acceptance over the dynamic energy 
range of the telescope system. Thus the second $\gamma$-ray selection criterion was 
$<\tilde{w}>\, \leq 1.2$, which accepts most of the $\gamma$-ray showers ($\sim 96$\%) 
while the corresponding acceptance of cosmic ray showers is 20\%. \\
In stereoscopic observations the impact distance of the shower axis to a system 
telescope can be measured with an accuracy $\leq 10$ m. The energy of a $\gamma$-ray 
shower is defined by interpolation over the ``size'' parameter $\rm S$ (total number of 
photoelectrons in \v{C}erenkov light image) at the fixed impact distance $R$, as 
$\rm E = f_{MC}( S,R,\theta)$, where $\theta$ is the zenith angle and $\rm f_{MC}$ 
is a function obtained from Monte Carlo simulations. The rms error of the energy 
determination is $\rm \Delta E/E \sim0.18$. The energy distribution for the ON- and 
OFF-source events, after the orientation and shape image cuts, was histogramed over the 
energy range from 500 GeV to 30 TeV with 10 bins per decade. The $\gamma$-ray energy 
spectrum was obtained by subtracting ON- and OFF-histograms and dividing the resulting 
energy distribution by the corresponding collection area and the $\gamma$-ray acceptance. \\
The collection area for $\gamma$-rays rises very quickly in the energy range near the energy 
threshold of the telescope system, which is 500 GeV, whereas it is almost constant at the 
energy $\geq 3$ TeV. Even slight variations of the trigger threshold could lead to noticeable 
systematic changes in the predicted spectral behaviour in the energy range of 
$\sim 0.5-1$ TeV. In fact the trigger threshold is washed away around the norminal value 
of 10 (8) ph.e. This effect leads to a noticeable probability for ``sub-threshold'' triggers. In 
addition, the trigger level for different camera pixels is slightly different even 
after very accurate adjustment of the high voltage using the calibration laser runs. 
Measurements of the trigger setting for a number of camera pixels revealed the variations in 
the trigger threshold of $\sim 10$\%. These variations were implemented into simulations 
in order to estimate the corresponding systematic error of the energy 
spectrum at low energies, $\leq 1$ TeV.   \\
The unique $\gamma$-ray statistics from the Mrk 501 allows the measurement of the energy 
spectrum up to a few tens of TeV. However, detection of \v{C}erenkov light images with 
extremely large amplitudes - several thousands of ph.e. - is complicated by the  
nonlinearity in the PMT response as well as by saturation in the 8 bit Flash-ADC readout. 
Measurements of the photomultiplier response under the high light loading over the 
extended sample of the EMI 9073 PMTs provided us with a calibration function used to 
correct the image amplitudes. The readout of the HEGRA IACT is based on the sampling of  
\v{C}erenkov light time impulse by the 16 FADC bins of $\sim 8$ ns each (Hess et al., 
1998). The time pulses from the air showers with a full width at half maximum of 
a few ns were widened using an electronic scheme in order to fit into several 
FADC bins for the accurate measurement of the time profile. The smoothing of the FADC signal 
was unfolded back to the impulse, which almost always fits 2 FADC bins. The calibrated 
amplitude, summed over two FADC bins, is used as a measure of the pixel signal. For the 
high energy air showers the FADC signals run into saturation and the simple unfolding 
procedure fails. For such pulses the initial amplitude is reconstructed using 
the additional calibration function obtained by the simultaneous measurements of light 
flashes with FADCs and a 14 bit ADC. This procedure drastically extends the dynamic range 
of the FADC readout. \\
To avoid the saturation problem one might only use images detected from air showers 
at large impact distances from the telescope system (e.g. beyond 150 m). The size 
of these images is very small even for high energy events because of the low 
\v{C}erenkov light density far off the shower axis. However these images are very 
often truncated by the camera edge and do not allow a proper reconstruction of the 
shower impact point and the shower energy. This effect becomes less important in 
observations at large zenith angles because of the high shower maximum height (the images 
shrink to the camera center). The increase of the impact distance limit from 200 m up to 
250 m for a zenith angle of 30 degree gives an increase in the $\gamma$-ray rate of 
$\sim$20-30 \% whereas it introduces large systematic errors due to the inferior energy 
reconstruction. In the present analysis we applied a restriction on the 
shower impact distance from the central telescope of the system at 200 m. 
In addition we performed a large number of consistency and systematic checks. We analysed separately 
2-fold and 3-fold coincidence events to reproduce the same spectral shape. The different 
orientation and shape cuts were used in order to check how robust the energy 
reconstruction procedure is. Finally the different approaches converged to the 
resulting energy spectrum shape. 

\begin{figure}[htb]
\begin{center}
\includegraphics[width=0.7\linewidth]{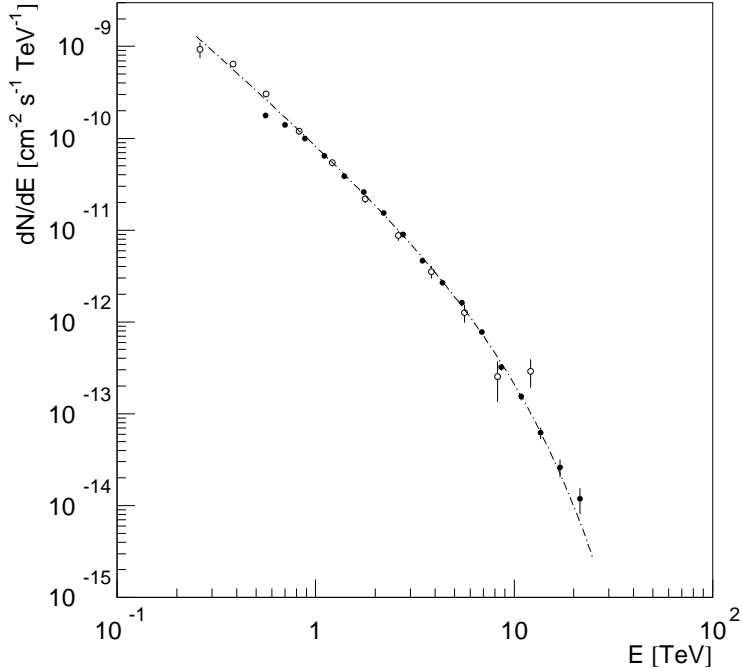}
\caption{\protect \small The energy spectrum of Mkn 501 (preliminary) as measured by the 
HEGRA IACT array (filled circles). 
The combined power law plus exponent fit of the
HEGRA data is shown by dotted-dashed curve. The Mkn 501 spectrum measured by the
Whipple group (open circles) is taken from Samuelson et al, 1998.}
\end{center}
\end{figure}
     
\noindent
{\bf 4. Results}

The Mkn 501 $\gamma$-ray flux, as averaged over the entire observation period, was about 
three times the Crab flux. The daily averaged $\gamma$-ray rate showed strong variations with a 
maximum of 10 Crab detected on 26/27 June 1997. As remarked in Aharonian et al., 
1997 the hardness ratio of the steepening Mkn 501 spectrum appeared to be independent of the absolute flux.  
The high $\gamma$-ray detection rate provided event statistics of a few hundreds within 1 
day's observations ($\sim 3-5$ hours) which is enough for the energy spectrum evaluation in 
the energy range 1-10 TeV. The data analysis of the energy spectral shape on a daily 
basis did not reveal any substantial correlations between the $\gamma$-ray flux and the 
spectral behaviour (Aharonian et al., 1998b). This justified the measurement of a 
time-averaged energy spectrum of Mkn 501 in the high state of its activity. \\
The time averaged spectrum of Mkn 501 over the entire energy range from 500 GeV to 24 TeV is shown 
in Figure 1. The vertical error bars correspond to the statistical errors. Note that  
the systematic errors at energies below 1 TeV appear to be quite large, and reach $\sim$50 \% at 500 GeV. \\
The Mkn 501 energy spectrum shows a {\it gradual steepening} over the entire energy range. 
The energy bin of 19-24 TeV contains an excess of 3.3$\sigma$ significance 
(23 $\gamma$-rays against 13 cosmic rays). The steep energy spectrum plus 20 \% energy 
resolution still allow these 23 $\gamma$-rays to be spilled out from lower energy bins. 
Therefore the energy spectrum is consistent with the hypothesis of a maximum energy for the 
detected $\gamma$-rays of 18.5 TeV at the $2\sigma$ confidence level. The shape of the 
energy spectrum is well determined by a power law fit  with an exponential cutoff. A fit of 
the data over the energy region where the systematic errors are small, i.e. from 1.25 TeV to 24 TeV, 
gives 
$$dN/dE \, = \, 9.7\, \pm 0.3\, (\textrm{stat})\, \pm 2.0\, (\textrm{syst}) 
\cdot 10^{-11}\, E^{-1.9 \, \pm 0.05 \,(\textrm{stat})\, \pm 0.05 \,(\textrm{syst})}$$
\begin{equation}
\exp\left[-E/(5.7 \,\pm  1.1 (\textrm{stat}) \,\pm 0.6 \,(\textrm{syst})
{\rm TeV})
\right] \,\,
{\rm [cm^{-2} s^{-1} TeV^{-1}]}.
\end{equation}
The logarithmic slope of the energy spectrum (``power law index'') is 
$1.8 \pm 0.14 \pm 0.82$ in the energy range 1.25-5.0 TeV, and $3.7\pm 0.33 \pm 0.6$ 
above 5 TeV. 

\noindent
{\bf 5. Discussion}

Detection of TeV $\gamma$-rays from Mkn 501 leads to the conclusion that they are 
produced within the relativistic jet. If not, the $\gamma$-rays will be absorbed inside 
the source due to the $\gamma$-$\gamma$ interactions with the soft radiation field.
This constrains the Doppler factor of the jet $\delta \geq 10$. Assuming that the 
observed spectrum of Mkn 501 is a power law modified by the source internal 
$\gamma$-$\gamma$ absorption, one can get more accurate determination of the Doppler 
factor of the jet $\delta \simeq 8.5$. However, since there is a number of reasons 
for the steepening of the TeV spectrum, this estimate can only be considered as a 
lower limit on the Doppler factor of the relativistic jet. \\
The curvature of Mkn 501 energy spectrum could occur due to a several physics processes
related to the blazar mechanism of the VHE $\gamma$-ray emission, e.g., the exponential 
cutoff in the spectrum of accelerated particles, as well as the energy spectrum 
modifications caused by $\gamma$-ray propagation in the intergalactic medium. Thus the 
extragalactic background light attenuate the intrinsic blazar $\gamma$-ray energy 
spectrum substantially at the high energy end via the pair-production interactions of 
the $\gamma$-rays with the diffuse background photons.
The extension of Mkn 501 spectrum beyond 10~TeV gives a strong upper limit on the 
intergalactic IR field in a region between 2 and $\sim 40$ $\mu \rm m$ as $\nu F(\nu ) 
\sim 4 \,\, \rm nW m^{-2} sr^{-1}$ and corresponding optical depth at 18 TeV, $\tau \leq 3$. 
Since this upper limit is rather close to the current
theoretical as well as phenomenological estimates of the intergalactic radiation field, 
it would be even possible to derive the absolute fluxes based on the TeV energy spectrum.
However the current uncertainties in the spectral energy distribution of the intergalactic
radiation does not allow us to reconstruct definitely the intrinsic energy spectrum of 
Mkn 501. \\
We believe that the real progress in understanding of the TeV $\gamma$-ray emission 
can be achieved by the correlated spectral and temporal studies of the 
BL Lac objects in the X-ray and TeV $\gamma$-rays obtained during the multiwavelength 
campaigns as well as by the future detections of the blazars at the different states of 
activity, and at the different distances up to several hundreds Mpc.

\noindent
{\bf Acknowledgements}

{\small 
The support of the German Ministery for Research and Technology BMBF and of the Spanish 
Research Council CICYT is gratefully acknowledged. We thank the Instituto de Astrof\'{i}sica 
de Canarias for the use of the site and providing excellent working conditions. We 
gratefully acknowledge the technical support staff of Heidelberg, Kiel, Munich, and 
Yerevan.} 

\noindent
{\bf References}

{\small 

\noindent
Aharonian F., et al., 1997, {\it Proc. 4th Compton Sym.}, Part 1, Williamsburg, 1631 \\
Aharonian F., et al., 1998a, {\it A\&A}, 327, L5-L8 \\
Aharonian F., et al., 1998b, submitted to {\it A\&A} \\  
Daum A., et al., 1998, {\it Astropar. Physics}, vol. 8, 1-2, 1 \\
Hess M., et al., 1998, {\it Astropar. Physics.}, in press \\
Konopelko A., et al., 1998a, {\it Astropar. Physics}, in press \\
Konopelko A., et al., 1998b, {\it Rayos cosmicos 98: Proc. 16th ECRS}, Ed. J.Medina, 
Universidad de Alcala,  523 \\
Lorenz E., et al., 1998, {\it Astropar. Physics} in press \\
Mirzoyan R.,  et al., 1994, {\it NIM,} A 351, 513 \\
Reynolds P., 1991, {\it Proc. 22nd ICRC, Dublin}, 1, 496 \\
Samuelson F., et al., 1998, {ApJ}, 501, L17  \\
Weekes T., et al., 1997, {\it Proc. 4th Compton Sym.}, Part 1, Williamsburg, 361 
}

}
\end{document}